\documentclass[a4paper,12pt]{article}
\pdfoutput=1 
             
\usepackage{jinstpub} 
\usepackage[utf8x]{inputenc}
\usepackage[version=3]{mhchem}
\DeclareMathOperator\erf{erfc}
\usepackage{textcomp}
\usepackage{multirow}
\title{Enhancement of the X-Arapuca photon detection device for the DUNE experiment}
\author[a,b] {C.~Brizzolari,} 
\author[c,d] {S.~Brovelli,}
\author[c,d] {F.~Bruni,}
\author[b,a] {P.~ Carniti,}
\author[b,a,1] {C.~M.~Cattadori,\note{Corresponding author.}}
\author[a,b] {A.~Falcone,}
\author[b,a] {C.~Gotti,}
\author[e] {A.~Machado,}
\author[c,d] {F.~Meinardi,}
\author[b,a] {G.~Pessina,}
\author[e] {E.~Segreto,}
\author[e,b] {H.~V.~Souza,}
\author[a,b]{M.~Spanu,}
\author[a,b] {F.~Terranova,}
\author[a,b] {M.~Torti}

\emailAdd{carla.cattadori@lngs.infn.it}
\date{Draft 1 (15th April 2021)}

\affiliation[a]{University of Milano Bicocca, Physic Department, Piazza della Scienza 3, Milano - Italy}
\affiliation[b]{INFN Milano Bicocca, Piazza della Scienza 3, Milano - Italy}
\affiliation[c]{University of Milano Bicocca, Department of Materials Science , Via Cozzi 55, Milano - Italy}
\affiliation[d]{Glass to Power s.p.a., Via Fortunato Zeni 8 - Rovereto - Italy}
\affiliation[e]{Instituto de Física Gleb Wataghin, Universidade Estadual de Campinas - Unicamp, \\ Rua Sérgio Buarque de Holanda, No 777, CEP 13083-859 Campinas, SP, Brazil}

\abstract{In the Deep Underground Neutrino Experiment (DUNE), the VUV LAr luminescence is collected by light trap devices named X-Arapuca, sizing $\sim$~(480~$\times$~93) mm$^2$.
Six thousand of these units will be deployed in the first DUNE ten kiloton far detector module. In this work we present the first characterisation of the photon detection efficiency of an \mbox{X-Arapuca} device sizing $\sim\,$(200~$\times$~75)~mm$^2$ via a complete and accurate set of measurements along the cell longitudinal axis with a movable $^{241}$Am source. The MPPCs photosensors are readout by a cryogenic trans-impedance amplifier to enhance the single photoelectron sensitivity and improve the signal-to-noise while ganging 8 MPPC for a total surface of 288~mm$^2$. Moreover we developed a new photon downshifting polymeric material, by which the X-Arapuca photon detection efficiency was enhanced of  about +50\% with respect to the baseline off-shell product deployed in the standard device configuration. The achieved results are compared to previous measurements on a half size X-Arapuca device, with a fixed source facing the center, with no cold amplification stage, and discussed in view of the DUNE full size optical cell construction for both the horizontal and the vertical drift configurations of the DUNE TPC design and in view of liquid Argon doping by ppms of Xe. Other particle physics projects adopting Liquid Argon as target or active veto, as Dark Side and LEGEND or the DUNE Near Detector will take advantage of this novel wavelength shifting material.}

\begin{document}
\maketitle
\flushbottom

\newpage 
\section{Introduction}
\label{sec:intro}

Liquid Argon (LAr) is a dielectric material featuring high density (1.4 g/cm$^3$), moderate radiation length (14 cm), and high electron mobility ($\sim$ 550~$\text{cm}^2\cdot(\text{V}\cdot\text{s})^{-1}$); thanks to these and other features, in 1977 the LAr Time Projection Chamber (TPC) technology was proposed by C. Rubbia as a superior alternative to bubble chambers for highly granular tracking and calorimetric detectors. ICARUS~\cite{ICARUS} showed the feasibility of large LAr based TPC and its first-grade calorimetry and imaging features. Since then, thanks to the excellent LAr light yield of 51000 photons/MeV,  the LAr TPC design further developed integrating the LAr luminescence pulse detection: the particle energy losses populate singlet (S) and triplet (T) states of Ar dimers (Ar${_2}{^*}$), that de-excite with characteristic time of 6 ns ($\tau_S$) and $\sim$1.4 $\mu$s ($\tau_T$) respectively, emitting 128 nm photons. The detection of the VUV LAr luminescence light was first made possible by the development of cryogenic photomultiplier (PMT) and of the technique to coat PMT entrance windows and detector reflective surfaces with 200~-~500~$\mu$g/cm$^2$ organic dyes as TPB or pTP\footnote{pTP is para-Terphenyl, TPB is Tetra-phenyl Butadiene} to effectively downshift the 128 nm photons to longer wavelength. In the last 20 years, the LAr light detection was further boosted by the MPPC or Silicon Photomultiplier (SiPM) technology that substituted PMTs. 

The GERDA project~\cite{gerda_phase2} successfully adopted Linde LAr 5.0 grade as a radiopure vetoing active shield and cooling agent of Ge detectors: the LAr light was captured by  quartz fiber coupled to SiPMs. Dark Side~\cite{darksidecollaboration2021separating} choose fossile Argon as a monolithic ultra pure and radiopure homogeneous dark matter target, depleted in the cosmogenic $^{39}$Ar and $^{42}$Ar isotopes.

The DUNE project aims to probe neutrino CP violation and mass hierarchy, and to characterize in term of energy, flavour and arrival time the neutrino burst profile emitted in the novae explosions, exposing at least two 17.5-kton far detectors equipped with both charge and LAr light detection to the LBNF neutrino beam. To fulfill the supernovae neutrino scientific program, DUNE requires an average light yield of >20 photoelectron/MeV with a minimum >0.5 photoelectron/MeV, corresponding to photon detection efficiency (PDE) of 2.6\% and 1.3\%~\cite{dune_techrep}, respectively. In the DUNE far detector (FD), 36000 units of light detectors named X-Arapuca will be deployed at the level of the anode planes. The optimization of its PDE is a first priority to secure the DUNE scientific program. 

This work describes and discusses accurate X-Arapuca PDE measurements in LAr and shows how it was improved of about +50\%.

The Vertical Drift design recently proposed for the second DUNE FD module will take advantage of this achievement to increase the size of the photodetector basic unit while keeping the same number of SiPMs, hence reducing the coverage. 

Two X-Arapuca devices of the same size, shape and SiPMs coverage of those tested in the present work have been deployed in year 2020 in the protoDUNE detector, that along about 10 months extensively tested  on a large scale the effect of increasing concentrations of Xe in LAr, ranging from 1  to $\sim$20 ppm on the detected light pulse profile. 
Xe doping and the Power over Fiber technique to power and readout the photosensors, are  presently considered an option for the second FD DUNE module, where the charge will be vertically drifted 
The know how to develop and produce wavelength shifting (WLS) slabs sizing 60~$\times$~60~cm with attenuation lengths of $\mathcal{O}$(m), and of methods to measure and optimize the PDE efficiency of such large area  units w.r.t. the wavelength of the incoming radiation and of the photosensor coverage will be beneficial for the DUNE Photon Detector. 

\section{The X-Arapuca device}
\label{sec:xarapuca}
%

\begin{figure}[htbp]
\center
\includegraphics[width=0.3\textwidth]{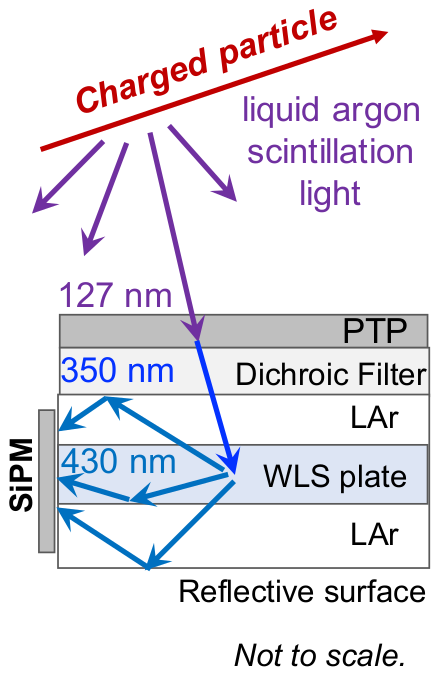}
\caption{The X-Arapuca concept scheme~\cite{dune_techrep}.}
\label{fig:xascheme}
\end{figure}

The X-Arapuca (XA) light trap is an evolution of the Arapuca design~\cite{Machado_2018,arapuca_proposal}; it is a reflective box equipped with an entrance window, two photon downshifting stages, one dichroic filter and one light guide coupled to SiPMs (see Fig.~\ref{fig:xascheme}). 

As illustrated in Fig.~\ref{fig:xascheme}, the pTP coating ($\sim$ 550 $\mu$g/cm$^2$) of the device entrance window efficiently (>95\%) downshifts photons from 128 nm to 350 nm~\cite{pTP}, isotropically re-emitting about 50\% inside the trap that is flooded by LAr: those absorbed by the wavelength shifting~(WLS) guide 
are again isotropically re-emitted at 440 nm and if above the critical angle are driven to the SiPMs, located at the WLS slab edges, else they leave. The dichroic filter (cutoff at 400 nm), deposited on the inner side of the entrance window, and the 3M Vikuiti\textsuperscript{\textregistered} reflector, lining the bottom and the sides of the light trap, bounce them back and forth until they are finally either detected by SiPMs or absorbed by materials i.e. lost.

The WLS slab is coupled to 4 groups of 4 SiPMs each produced by Hamamatsu, model S14160-6050HS (6~$\times$~6) mm$^2$, 50 $\mu$m pitch, peak sensitivity wavelength ($\lambda_{p}$) 450 nm, photon detection probability at $\lambda_{p}$ $\sim$50\%, terminal capacitance of 2.0 nF, and operated at +2.7 overvoltage above the breakdown~\cite{sipms_DS}, that at LAr temperature is found to be at 33.0~$\pm$~0.1~V. The sixteen SiPMs are non selected and the bias is unique for all of them.

\begin{figure}[htbp]
\center
\includegraphics[width=0.28\textwidth]{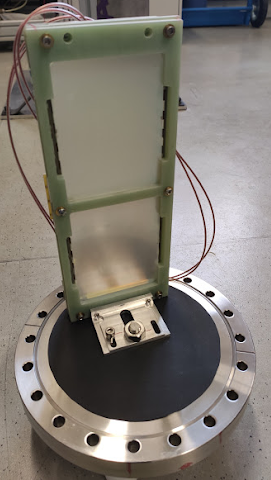}
\includegraphics[width=0.30\textwidth]{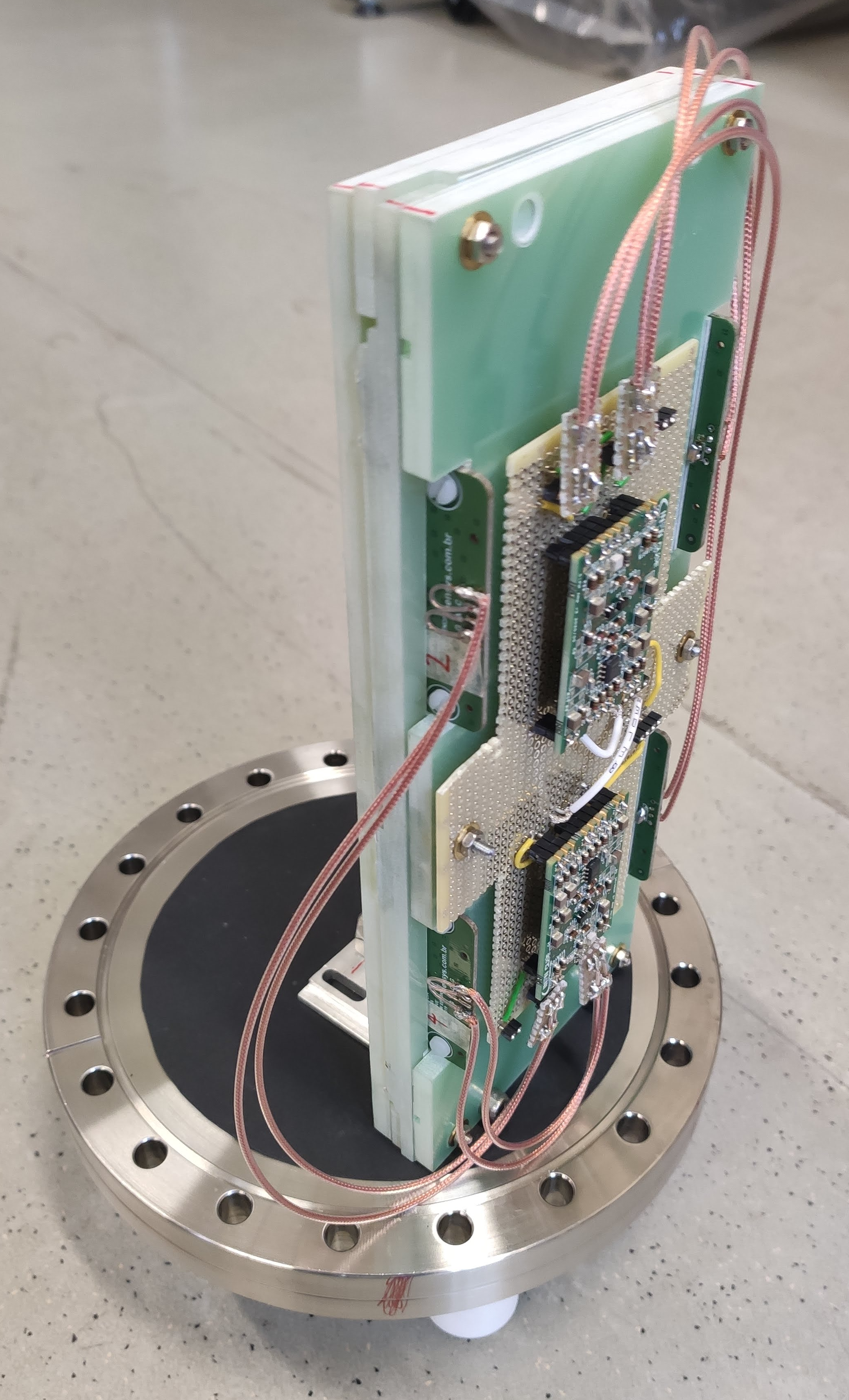}
\caption{Front and back views of the X-Arapuca device mounted on the bottom DN150 flange of the cryogenic chamber. On the back two readout transimpedance amplifier circuits are mounted on a service board providing the input/output electrical connections and contacts to the SiPMs. Eight SiPMs are ganged in input to each circuit.}
\label{fig:XAonFlange}
\end{figure}

The XA concept has been declined in several sizes: one window size~\cite{first_lar_test} (100~$\times$~75)~mm$^2$, two windows size (200~$\times$~75)~mm$^2$, that will be deployed in SBND~\cite{sbnd}, and 6 windows size (490~$\times$~93)~mm$^2$, which is the basic photon detection unit for the DUNE far detector module~\cite{dune_techrep}. In the present work we characterize the two window unit that is shown in Fig.~\ref{fig:XAonFlange}.

\section{The cold amplifier}
\label{sec:coldamp}

\begin{figure}[htbp]
\center
\includegraphics[width=0.7\textwidth]{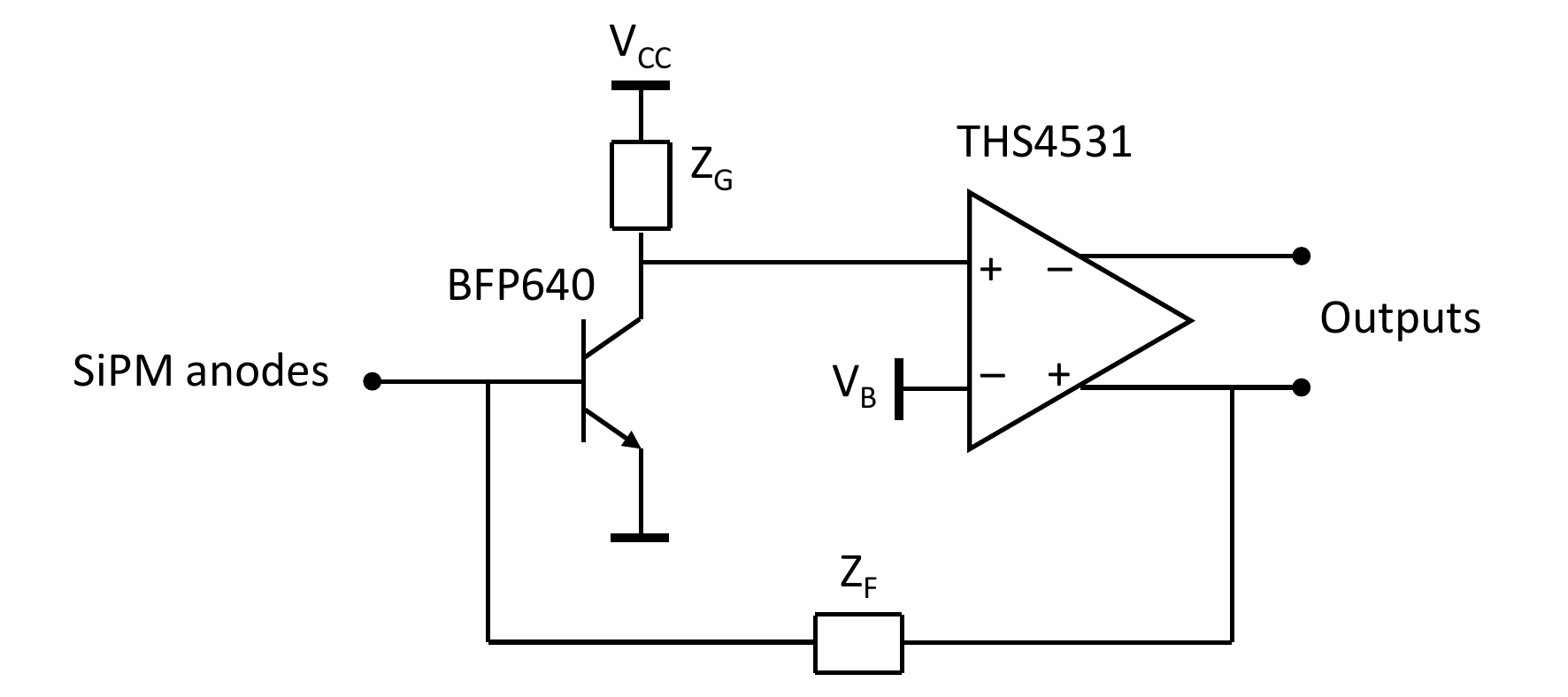}
\caption{Simplified schematic of the cold amplifier circuit.}
\label{fig:coldamp}
\end{figure}

The SiPM anodes are connected in parallel and read out by a transimpedance amplifier operated at cold.
Figure~\ref{fig:coldamp} shows a simplified schematic of the circuit, while Fig.~\ref{fig:XAonFlange}, right panel, shows  two cold circuit units mounted on the back of the XA device, connected to the SiPMs boards.
The input transistor is a BFP640 SiGe HBT from Infineon, chosen to give low voltage noise ($0.4\ \textrm{nV}/\sqrt{\textrm{Hz}}$) at a low bias current ($0.3\ \textrm{mA}$).
The first stage is followed by a THS4531 fully differential operational amplifier from Texas Instruments, that gives the bulk of the open loop gain, and provides differential outputs able to drive a differential signal pair.
The amplifier operates at a single power supply rail of $3.3\ \textrm{V}$ and consumes about $2\ \textrm{mW}$.
The design of the amplifier is described in deeper detail in~\cite{Carniti_2020}.

In this work, two amplifiers were used to read out eight $\textrm{6 $\times$ 6}\ \textrm{mm}^2$ SiPMs each, corresponding to an input capacitance of about $\textrm{15\ nF}$.
To save on the total number of electrical connections passing through the vacuum flange, only one of the two differential outputs was acquired for each of the two amplifiers.
The low voltage noise of the amplifier allows clear definition of single photoelectron peaks, even at low overvoltage.

\section{The novel secondary wavelength shifter}
\label{wlsnew}
The secondary wavelength shifters (WLS) are produced by bulk polymerization of methyl methacrylate (MMA) doped with the active dye, the 2,5-Bis(5-tert-butyl-benzoxazol-2-yl)thiophene (BBT), using a cell casting process. The casted slabs are about (50~$\times$~50)~cm$^2$ large with a thickness of 0.4~cm and are finally cured at 90~°C for 24h.

They are then laser-cut in order to achieve several WLS slabs of the proper dimensions, finally the slabs edges are polished for ensuring the best optical coupling with the photon detectors. 

\begin{figure}[htbp]
\center
\includegraphics[width=0.7\textwidth]{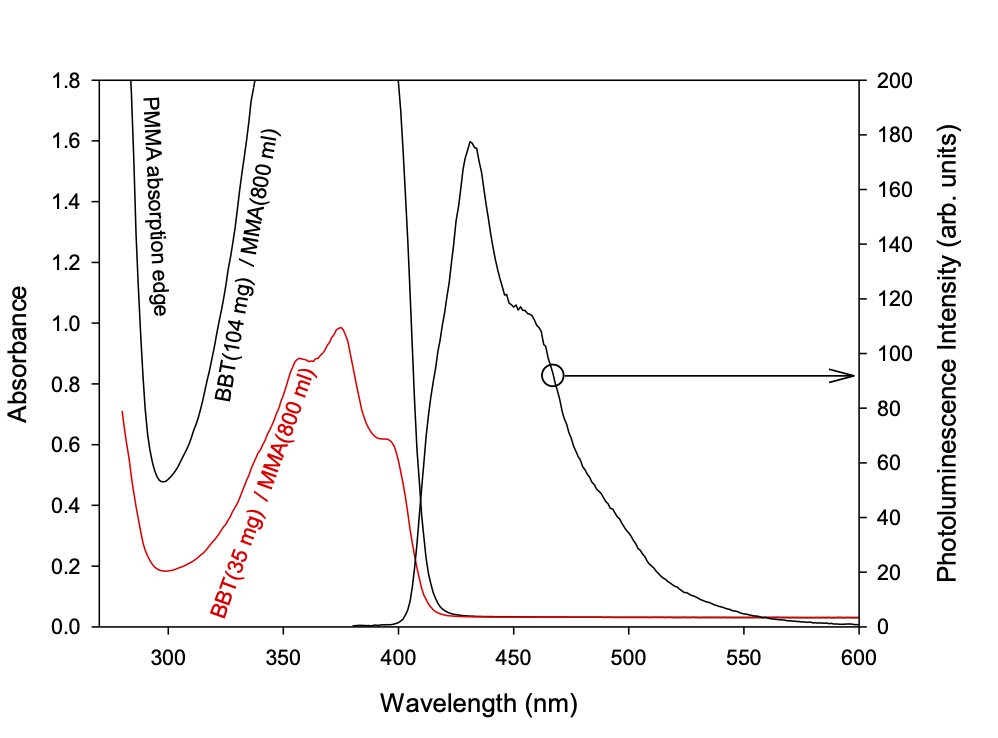}
\caption{The newly developed secondary wavelength shifters absorption and emission spectra, and the PMMA absorption.}
\label{fig:swls}
\end{figure}

The obtained Poly(methyl methacrylate) (PMMA), when undoped with the BBT, shows an absorption edge at around 300~nm thus resulting completely transparent to the emission of the primary WLS which is on the contrary perfectly resonant with the first absorption band of the BBT centered at 370~nm. Similarly, the emission of the BBT shown in Fig.~\ref{fig:swls}, centered at 430~nm, well matches the range of maximum efficiency of the SiPMs. The photoluminescence quantum yield (QY) of the BBT is reported to be quite large regardless the environment \cite{wls1}. In particular, in the PMMA matrix we measured a QY~>~90\%. 

It is important to note that the overlap between the BBT absorption and emission spectra is rather small which minimizes the reabsorption probability of the BBT photoluminescence.  This behavior, combined with the absorption of the PMMA which is below the detection threshold in this spectral range, implies a quite long attenuation distance of the BBT emitted light which can reach the detectors on the edges regardless of where it is generated.

The ideal concentration of dye depends on the balance between the need to absorb the radiation emitted by the primary WLS, which is enhanced by high concentrations of dye, and to minimize the reabsorption which, even if small, cannot be zero and increases by increasing the BBT amount. For this reason, we tested several secondary WLS with different BBT concentrations. For (21~$\times$~7.5)~cm$^2$ slabs the best result have been obtained by using 80 mg of BBT for every kg of PMMA. At this concentration more than 99\% of the primary WLS emission is absorbed while the reabsorption still remains negligible. In the following we will refer to these slabs as FB118. If necessary, small readjustments of the dye concentration can be easily done for re-optimizing the performances of secondary WLS with different sizes or with a different disposition/number of detectors on the edges.
The PMMA based WLS slabs have been tested to be cryo-resilient: following about 10 fast (cooling down in few minutes) thermal cycles from room to cryogenic environment, no cracks shown up.


\section{Room Temperature Measurements}
\label{sec:measurement_rt}

Room temperature measurements aim to assess both the photon detection uniformity (PDU) within each WLS slab 
and their relative photon detection efficiency (PDE), by means of a pulsed light source at 405 nm, hence the dichroic filter plate, was not deployed. 

The PDU of the X-Arapuca in its standard configuration, with the EJ-286 and the Vikuiti\textsuperscript{\textregistered}\footnote{An extended specular reflector with high $\sim$98\% reflectivity at 440 nm, extensively adopted in LAr detectors} applied on the X-Arapuca frame was first measured. Then the reflector was applied on the edges of the EJ-286 slab but the areas facing the SiPMs, and the measurement was repeated. The same procedure was adopted with the FB118.

As for the PDE, we performed  relative measurements of {\emph{i})} each bar with vs without the reflector on the edges, and {\emph{ii})} the FB118 (newly developed) vs. the EJ-286 (baseline) product.

The SiPMs were biased at 45 V (+3 OV) and read with four independent channels.

\subsection{Setup and methods}

\begin{figure}[htbp]
    \centering
    \includegraphics[width=.26\textwidth]{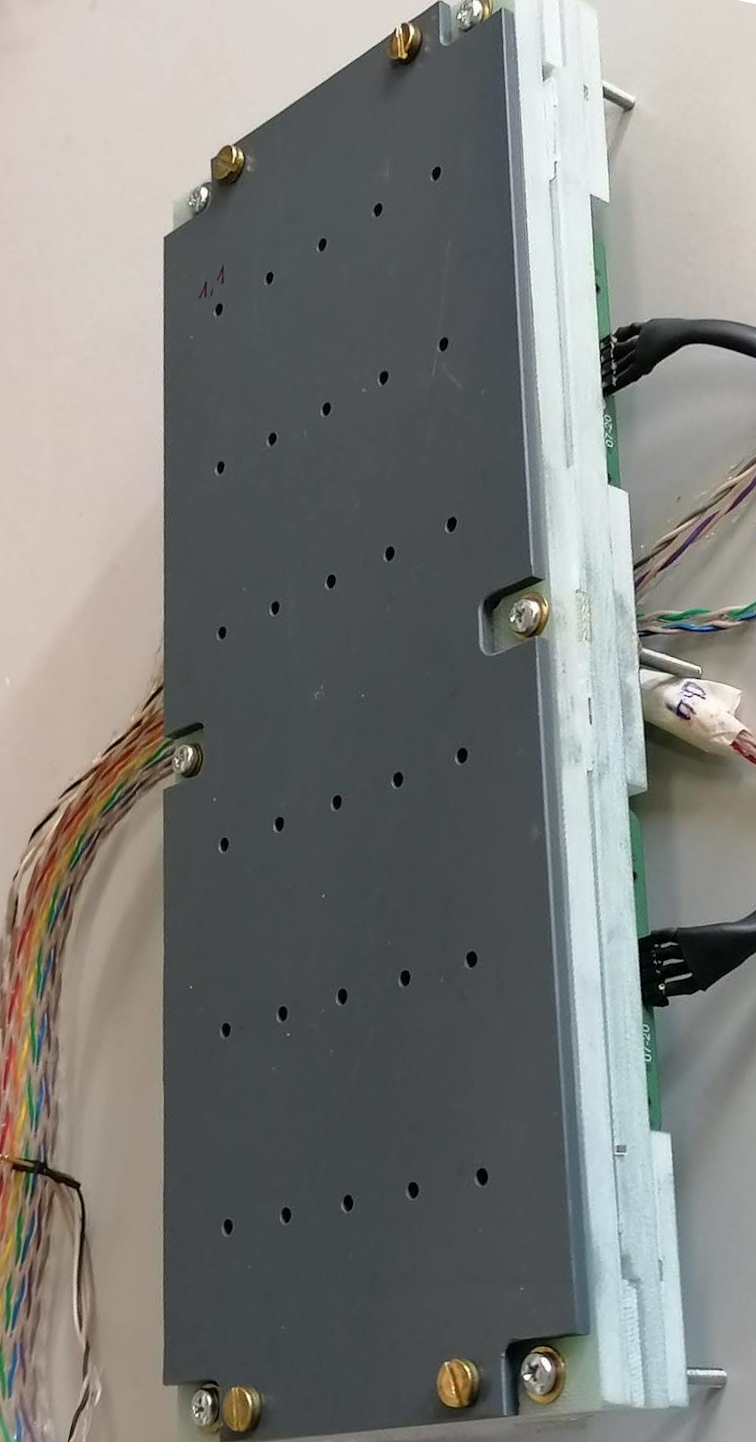}
    \includegraphics[width=.32\textwidth]{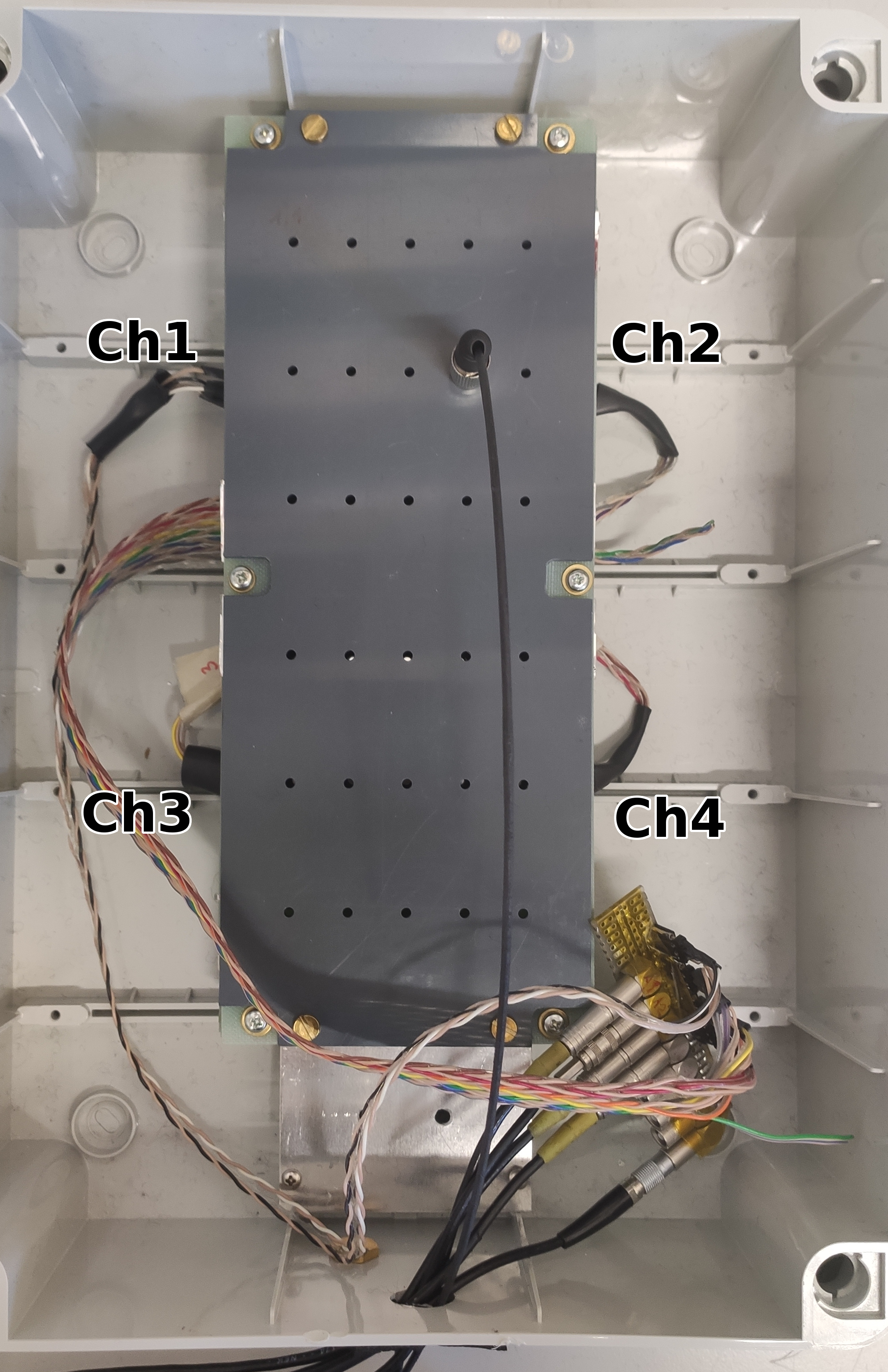}
    \caption{The X-ARAPUCA with the plastic mask and the optical fiber plugged.}
    \label{fig:xara_mask}
\end{figure}

The light from an LED pulsed source (CAEN, model SP5601, peak emission at 405~nm) was routed by an optical fiber to 30 equally spaced points of a mask located $\sim 5~\text{mm}$ above the top face of the scintillator slab. The mask ensures the reproducibility of the points position and the orthogonality of the light beam w.r.t. the WLS top surface (see Fig.~\ref{fig:xara_mask}). The assembly was operated inside a light-tight box.

\noindent The SiPMs signals were acquired with an oscilloscope (Rhode~\&~Schwartz~mod.~RTE1204); for each light entrance point the amplitude and the area of the four signals were registered. Regular acquisitions of the baseline (i.e.\ signal of the SiPMs dark current) were also performed to check the stability of the dark current. The net area of the light pulses is obtained by proper subtraction of the baseline from the area values.
From both the pulse amplitude and net area measurements the PDU matrix within each of the two WLS slabs is measured by the parameter $V_{ij}$ defined by Eq.~\ref{eq:RT_uniformity}, both with and without the Vikuiti\textsuperscript{\textregistered} directly applied on the edges.

\begin{equation}
\label{eq:RT_uniformity}
V_{ij} = \frac{V^{\mbox{ch1}}_{ij} + V^{\mbox{ch2}}_{ij} + V^{\mbox{ch3}}_{ij} + V^{\mbox{ch4}}_{ij}}{\mbox{Average}(V^{\mbox{ch1}} + V^{\mbox{ch2}} + V^{\mbox{ch3}} + V^{\mbox{ch4}})}
\end{equation}
The rms of the PDU matrix is then computed.

\begin{figure}[htbp]
    \centering
    \includegraphics[width=.39\textwidth]{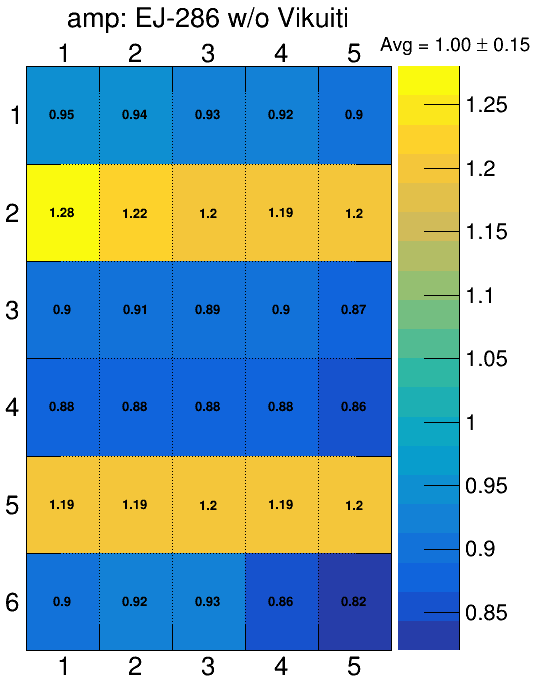}
    \includegraphics[width=.39\textwidth]{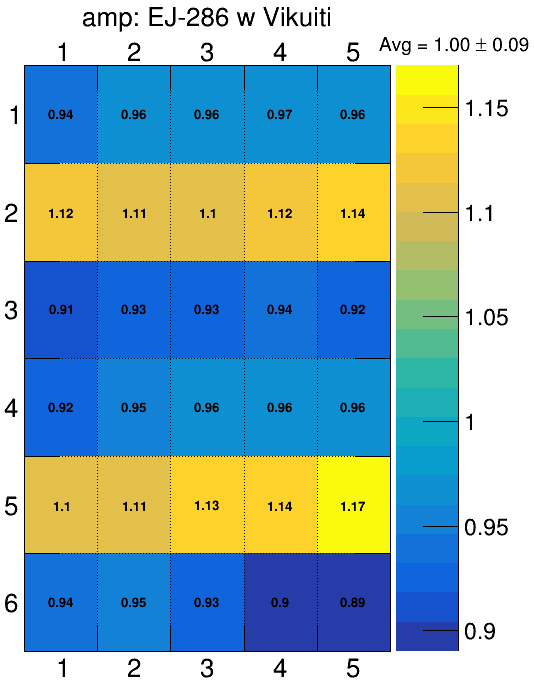}
    \caption{The PDU matrix of the EJ-286 slab, without (left) and with (right) Vikuiti\textsuperscript{\textregistered}.}
    \label{fig:ej_RT}
\end{figure}

\begin{figure}[htbp]
    \centering
    \includegraphics[width=.385\textwidth]{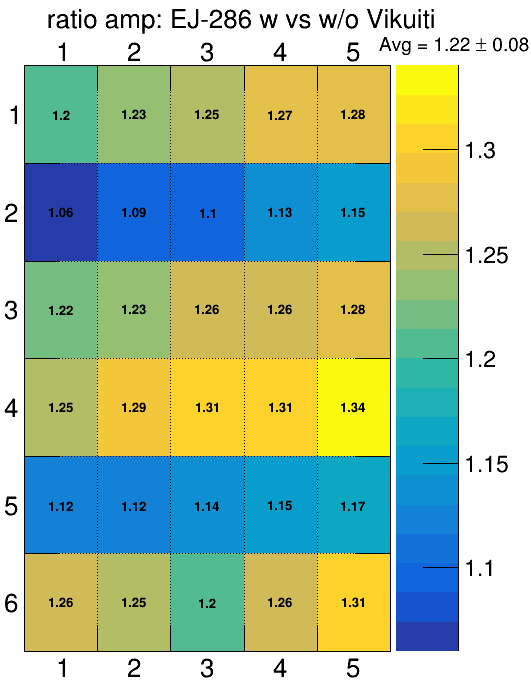}
    \includegraphics[width=.385\textwidth]{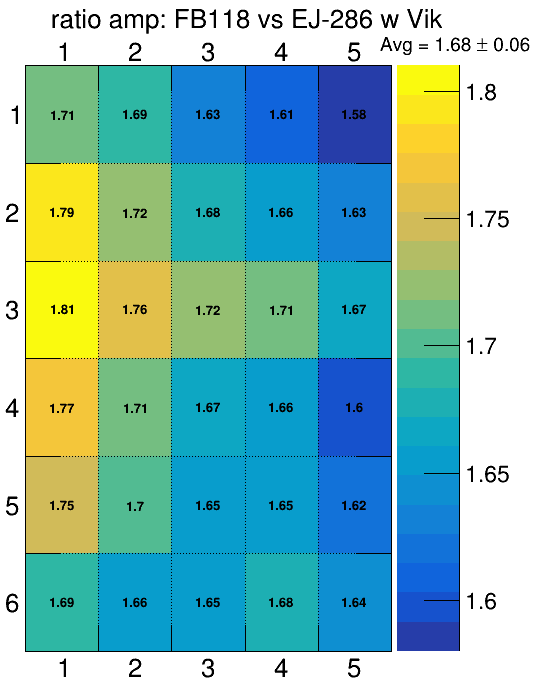}
    \caption{The PDE ratio matrix for the EJ-286 with/without Vikuiti\textsuperscript{\textregistered} (left), and for the FB118/EJ-286, both with Vikuiti\textsuperscript{\textregistered} (right). }
    \label{fig:ej_FB_RT}
\end{figure}

The PDE relative variation between the two slabs was then computed by the matrix of the ratios $R_{ij}$:

\begin{equation}
\label{eq:RT_ratio}
R_{ij} = \frac{A^{\mbox{ch1}}_{ij} + A^{\mbox{ch2}}_{ij} + A^{\mbox{ch3}}_{ij} + A^{\mbox{ch4}}_{ij}}{B^{\mbox{ch1}}_{ij} + B^{\mbox{ch2}}_{ij} + B^{\mbox{ch3}}_{ij} + B^{\mbox{ch4}}_{ij}}
\end{equation}

\noindent in which $A_{ij}$ and $B_{ij}$ is the signal collected in the $ij$-th point for the two samples, $A$ and $B$. Both the amplitude and the net area measurements gave compatible results. We found that when applying the Vikuiti\textsuperscript{\textregistered} reflector directly on the WLS bar edges the PDU rms decreases from 15\% to 9\% (Fig.~\ref{fig:ej_RT}) for the EJ-286 and from 17\% to 9\% for the FB118, and the PDE correspondingly increases of $\sim\,$22\% and $\sim\,$33\% respectively. This is illustrated in Fig.~\ref{fig:ej_FB_RT}~left for the EJ-286.


As for the FB118 PDE with respect to the EJ-286, an increase $G_\epsilon$ 
\begin{equation}
\label{eq:Gepsilon}
G_\epsilon = \left(\left<R^\text{FB118}_\text{EJ-286}\right>-1\right) \times 100 \; [\%],
\end{equation}
of ($\sim\,$50\%) $\sim\,$65\%  is observed when (not) applying the Vikuiti\textsuperscript{\textregistered}: the PDE ratio matrix is shown by Fig.~\ref{fig:ej_FB_RT}~right; for each position, the uncertainty is about of 0.2.




\section{Cryogenic measurements}
\label{sec:cryomeas}

\subsection{Setup and methods}
\label{sec:cryosetup}

\begin{figure}[htbp]
\center
\includegraphics[width=0.2\textwidth]{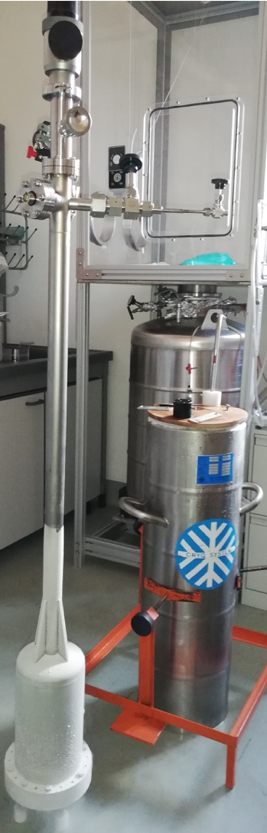}
\includegraphics[width=0.318\textwidth]{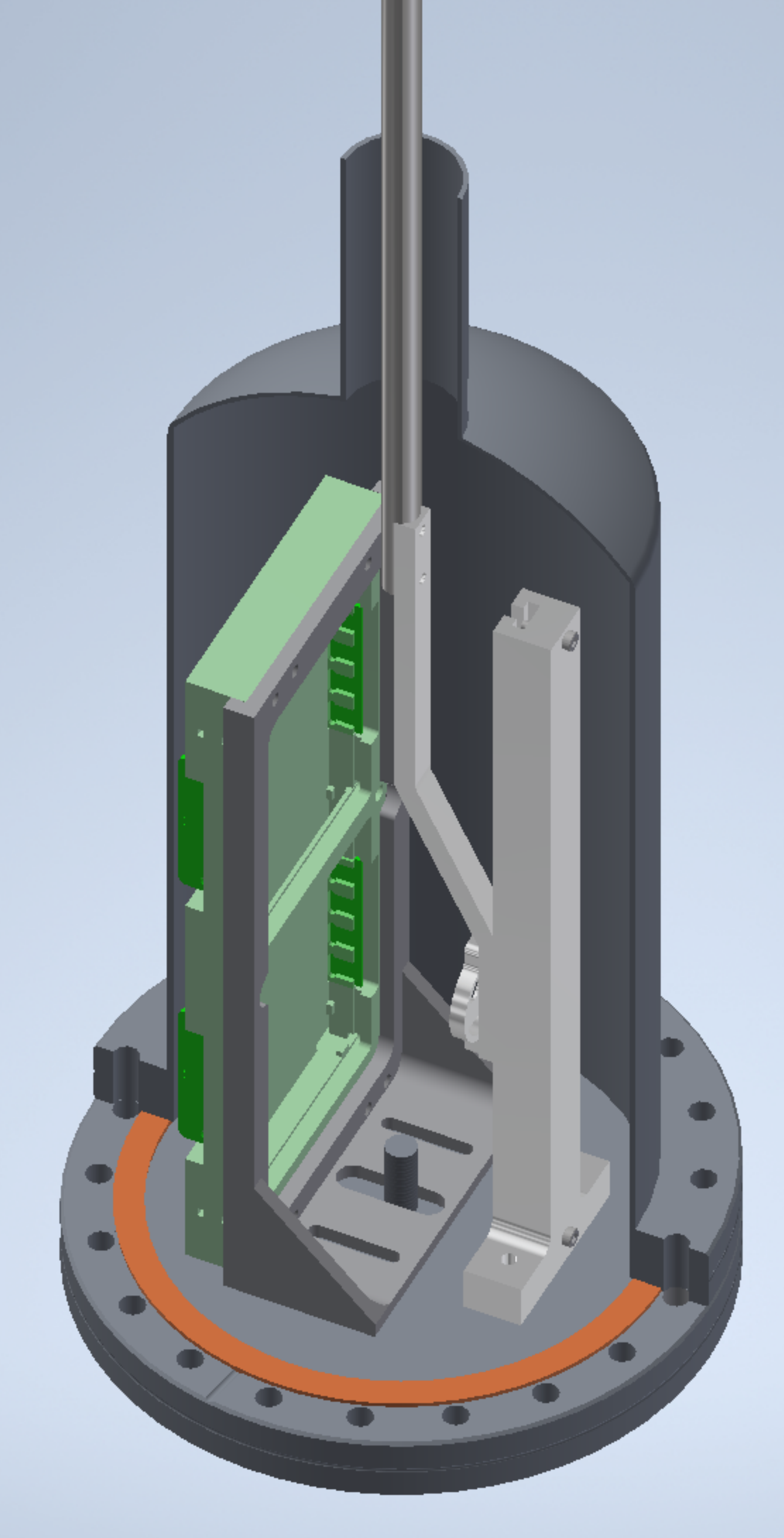}
\caption{The test chamber after removal from the LAr bath (left) and the 3D drawing of the chamber with the X-Arapuca and the source sliding rail.}
\label{fig:cryosetup}
\end{figure}

The setup shown in Fig.~\ref{fig:cryosetup} left hosted the X-Arapuca characterization in liquid Argon. A cryogenic stainless steel chamber sizing 250 mm diameter by 310 mm height, for a total volume of $\sim$5~liters is connected by a long steel pipe to the room temperature equipment (vacuum and gas lines, safety and vacuum/pressure sensors, electrical feedthroughs, a linear magnetic manipulator) mounted on CF flanges to the cryogenic environment. All the sensors and safety devices are stainless steel, whole-metallic, to minimize the outgassing in the chamber. The X-Arapuca named in the following \emph{device under test} (DUT) is located in the chamber, as shown by Fig.~\ref{fig:cryosetup} right. An exposed $^{241}$Am alpha source of 3.7 kBq activity can slide on an aluminum rail, thanks to the linear magnetic manipulator located in front of the DUT entrance windows at a distance of 55~$\pm$~1~mm along the whole device length. The source active surface is masked to $\sim$1/2 of the original surface to reduce the trigger rate. Moving the alpha source along the rail allows to scan the DUT along its central longitudinal axis.

Prior to the gas-Ar (GAr) liquefaction, the chamber is first pumped down to $\sim$10$^{-3}$~mbar, then the GAr  (6.0 grade) inlet is opened, while the chamber stays immersed in an external LAr bath. The pressure gradient generated by the liquefaction process, makes the GAr continuously flow from the bottle to the chamber where it is liquefied at the expenses of the evaporation of the external liquid Argon bath. The pressure in the chamber is regulated at about 1.4 bar, allowing to fill it in about 3 hours. A PT100 sensor on top of the DUT witnesses when it is fully immersed in LAr, and the GAr inlet is closed. By further refilling the external LAr bath, the setup offers up to 18 hours of stable data taking time, allowing to perform several alpha-scanning and cosmic rays data acquisitions.


\subsection{Measurements in Vacuum}
\label{sec:measurements_vacuum}

 To preliminary evaluate the photon downshifting and guiding efficiency of the two WLS slabs, we first worked in vacuum at cryogenic temperatures: the larger refraction index change between plastic and vacuum enhances the light guiding ($n_\text{WLS} \sim 1.58$) than in LAr ($n_\text{LAr} \sim 1.3$). 
 
 The alpha source was located a few millimeters from the pTP coating of the dichroic filter plate and the test chamber was pumped down to a pressure of about 10$^{-3}$ mbar. The alpha particles energy release in pTP arises its luminescence, peaked at 350~nm~\cite{pTP}. 

 The test chamber was slowly cooled down by an external Liquid Nitrogen bath: the PT100 sensor in contact with the DUT provides its actual temperature. The light pulses are calibrated by single photoelectron as described in Section \ref{sec:measurements_lar}. 

\begin{figure}[htbp]
    \centering
    \includegraphics[width=0.85\textwidth]{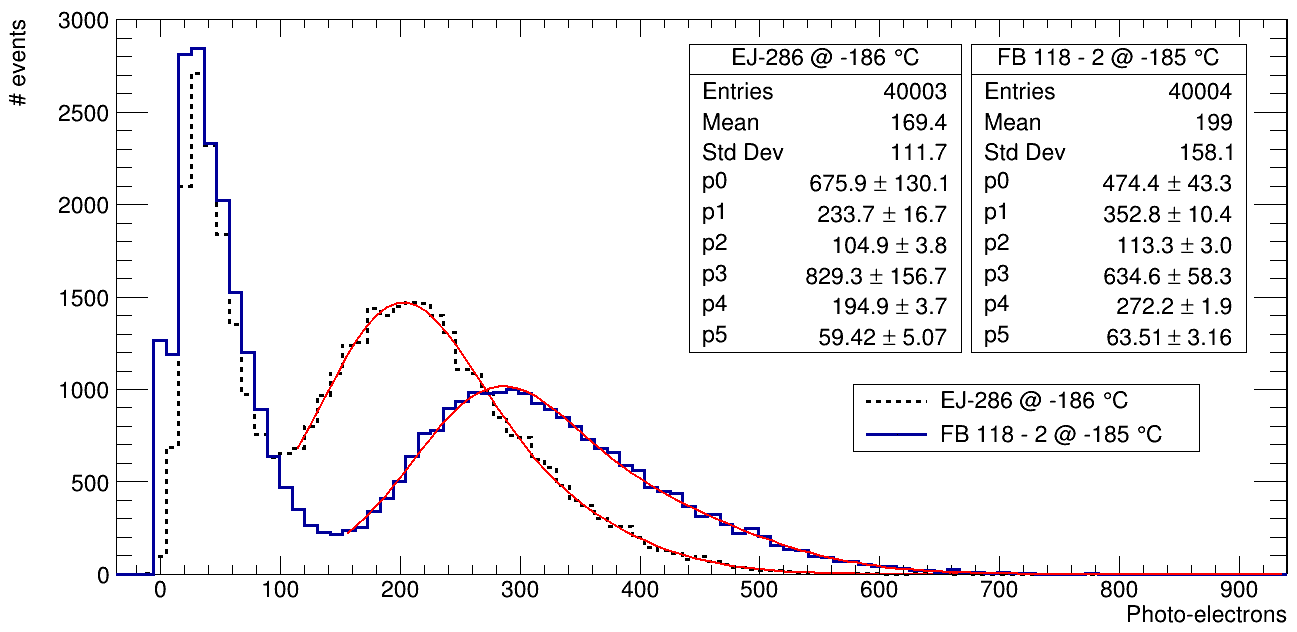}
    \caption{The $\alpha$ spectrum in photoelectrons, detected in LAr by the X-Arapuca with EJ-286 (black) and FB118 (blue). The fit with the sum of two gaussians is superimposed.}
    \label{fig:result_vacuum}
\end{figure}

Figure \ref{fig:result_vacuum} shows the calibrated $\alpha$ spectrum for the X-Arapuca equipped with each of the two WLS slabs operated at the same temperature. Here the SiPMs were directly coupled to the FADC input as the cold amplifier (see Sec.~\ref{sec:coldamp}) was not yet available, hence the uncertainty on the number of photo-electrons is about twice than in LAr tests, namely $\sim$5\%. To model the pTP light fraction ($\sim$50\%) emitted in the LAr and reflected back to the X-Arapuca by the alpha source holder, both spectra are fitted with the sum of two Gaussian's. Figure~\ref{fig:result_vacuum} reports the set of parameter from the fit: p0 and p3 correspond to the amplitudes, p1 and p4 to the means and p2 and p5 to the standard deviations of the two gaussian respectively. 

We measured a light collection increase of 42~$\pm$~13\% with the FB118 slab when comparing the peak position of the spectra. The peak position was derived from the fit: the quoted uncertainty includes both the s.p.e.\ calibration and the fit uncertainties. 

\begin{figure}[htbp]
    \centering
    \includegraphics[width=.85\textwidth]{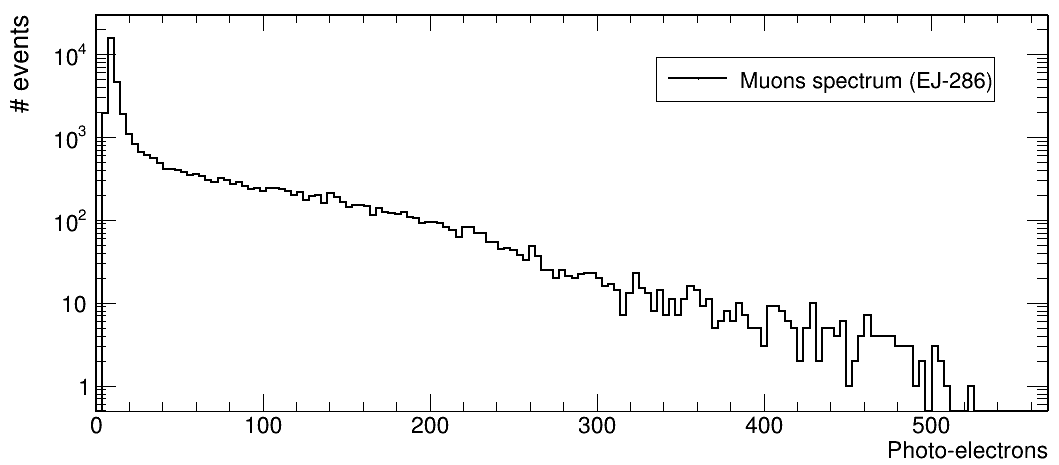}
    \caption{The muon spectrum for the X-Arapuca with EJ-286 in vacuum. 
    }
    \label{fig:muon_spectrum_vacuum}
\end{figure}

To investigate their intrinsic scintillation, the alpha source was removed and the X-Arapuca was exposed only to cosmic rays. 
Figure~\ref{fig:muon_spectrum_vacuum} shows the muon spectrum detected by the X-Arapuca with the EJ-286: the muons crossing the WLS slab arise its scintillation and a sizeable fraction of the spectrum has pulse light well above 100 photoelectrons. As a matter of fact, the EJ-286 WLS is based on polivyniltoluene (PVT), that, as well as Polystyrene, is known to be a scintillator~\cite{plasticsci14} with light yield of $\mathcal{O}(10^3)$ photons/MeV: the Eljen producer, in the EJ-286 data sheet\footnote{\url{https://eljentechnology.com/products/wavelength-shifting-plastics/ej-280-ej-282-ej-284-ej-286}}, mentions the attempt to reduce the direct scintillation component.

On the contrary, no scintillation was detected in the FB118 vacuum muon run and in fact PMMA is not known to scintillate. 

The WLS slabs intrinsic scintillation was also tested at room temperature (see Section~\ref{sec:measurement_rt}), with the alpha source in direct contact  ($\sim$1 mm in air): the $^{241}$Am $\alpha$-particles will release their energy within about 40~$\mu$m in the plastic. Light pulses were visible when irradiating the EJ-286 
while no light pulse was detectable from the FB118. When the pTP coated window was in place and the $\alpha$ source at contact, the pTP scintillation light was detectable with both slabs.

These results further motivated the comparison of the two WLS light-guides in Liquid Argon.

\subsection{Measurements in Liquid Argon}
\label{sec:measurements_lar}

Three X-Arapuca configurations were investigated in liquid argon: \emph{i)} the standard one with the EJ-286 WLS and the Vikuiti\textsuperscript{\textregistered} reflector applied only on the trap frame; \emph{ii)} as \emph {i} but with the reflector lining the slab edges to maximize the photon trapping efficiency and the collection uniformity (see Sec. \ref{sec:measurement_rt}); \emph{iii)} as \emph {ii} but the EJ-286 was changed with the FB118 wavelength shifter 


The main results are the relative comparison of the optical cell performances with the FB118 with respect to the EJ-286, the measurement of the absolute PDE of the X-Arapuca in each of the three described configurations, the assessment of our liquefaction procedures via the measurements of the relevant parameters of the singlet and triplet Ar$^*_2$ states: the latter is crucial to compare the PDE and PDU from individual liquefaction process.  

\begin{figure}[htbp]
    \centering
    \includegraphics[width=.49\textwidth]{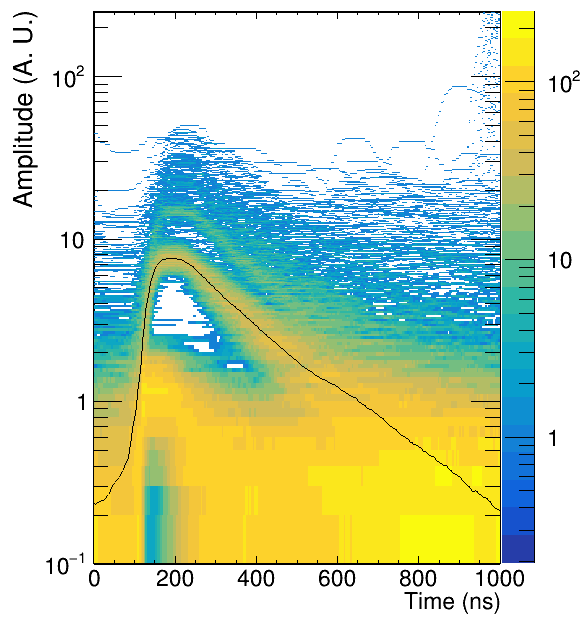}
    \includegraphics[width=.48\textwidth]{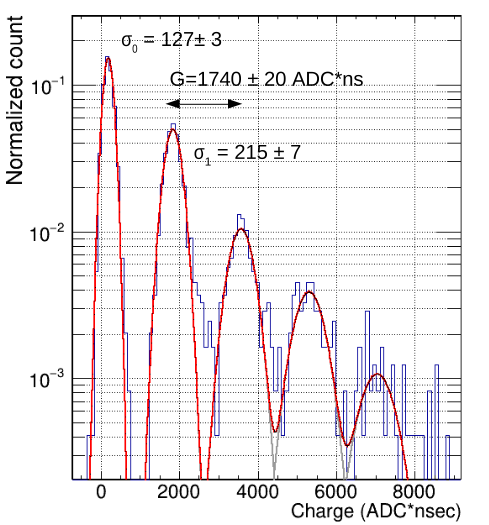}
    \caption{(Left) Persistence histogram of the selected waveforms together with the average waveform of one single photoelectron. (Right) Charge histogram of the selected peaks, the gain is defined by the distance between the first and second peak. The fit is described in the text.}
    \label{fig:sphe_calibration}
\end{figure}

The integrated light pulses are calibrated by mean of the single photoelectron (s.p.e.) integral. First the s.p.e. are searched in the waveforms pretrigger (the first 5 $\mu$s). All the pulses above 3.5~r.m.s  of the baseline are integrated for 0.46 $\mu$s and their trace ($\sim$1~$\mu$s from the onset) is stored. The de-noising algorithm  \cite{protoDUNE_first_results} is applied at the full-length waveforms while preserving the signal bandwidth \cite{denoising} . Figure~\ref{fig:sphe_calibration} left shows the 
\emph{envelope}-plot of the selected pulses, while Fig.~\ref{fig:sphe_calibration} right displays the histogram of the pulses charge, integrated over 0.46 $\mu$s. Five peaks are well visible: the noise and the one, two, three and four photoelectron peaks. The gain (G) of the SiPMs, later adopted to calibrate the $\alpha$ spectra is defined as the difference between the mean of the single and the two photoelectron peaks.

For each measurements, the gain and the S/N are derived, by fitting the spectrum with the sum of five Gaussians, to model the $n=0$ to 4 photoelectrons peaks. The fit function is shown in Fig.~\ref{fig:sphe_calibration} right: the $n=0$ noise and the single photoelectron peak ($n=1$) are fitted with two independent Gaussians\footnote{The mean of the noise is found to be slightly positive, as the peak searching algorithms bias it towards positive values. When not running the peak searching, the baseline mean is found at 0.026 ADC$\times$nsec with the same standard deviation.}. For the $n = 2$ peak the mean is free and the standard deviation is fixed at $\sqrt{n}\cdot\sigma_1$. For the $n>2$ peaks both the means and the standard deviations are fixed at $n\cdot G$ and $\sqrt{n}\cdot\sigma_1$ respectively. 

The average waveform of a s.p.e.\ is also shown in Fig.~\ref{fig:sphe_calibration} left, when selecting pulses whose integral charge is within one standard deviation (Fig.~\ref{fig:sphe_calibration} right).  
The single photoelectron calibration has an uncertainty of $\sim$2\% for each channel, with a maximum fluctuation of 4\% with respect to the lowest value, making it the largest source of uncertainty.

\begin{figure}[htbp]
    \centering
    \includegraphics[width=.46\textwidth]{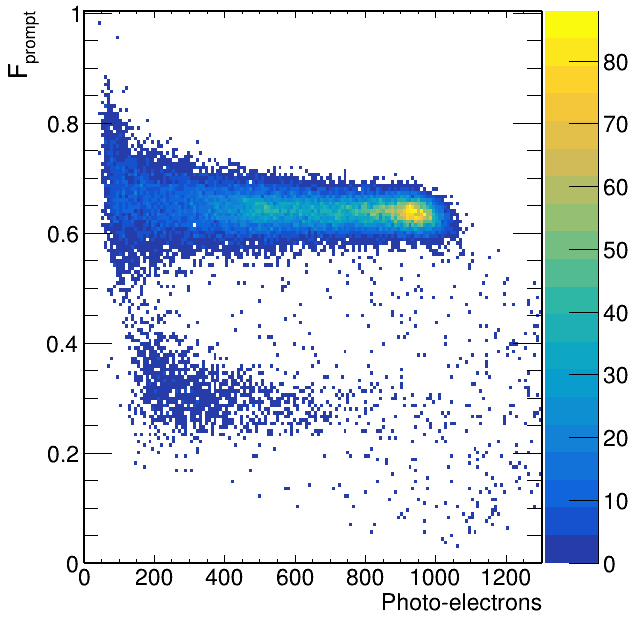}
    \includegraphics[width=.485\textwidth]{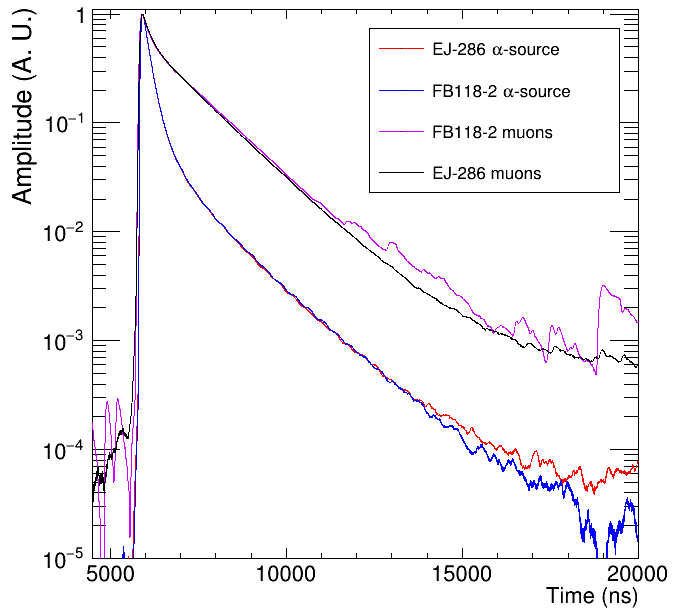}
    \caption{(Left) The events from an alpha run plotted in the plane F$_{\text{prompt}}$ versus number of p.e.:  $\alpha$ and muons have F$_{\text{prompt}}$>~0.5  and <~0.5 respectively.
    (Right) The normalized average waveforms of the events, selected on the F$_{\text{prompt}}$ classifier. 
    }
    \label{fig:averaged_waveforms}
\end{figure}

The LAr purity actually achieved in each measurement is measured by the triplet half-life for both alphas and muons. Particle identification is based on the F$_\text{prompt}$ classifier, defined as the ratio of the prompt (<600~ns) to the total charge~\cite{fprompt_psd,first_lar_test}. Figure~\ref{fig:averaged_waveforms} left shows the events in the plane F$_\text{prompt}$ versus total charge for an $\alpha$-run where cosmic muons also trigger. The events with F$_\text{prompt}$ > 0.5  are alphas, those < 0.5 are muons.

Figure~\ref{fig:averaged_waveforms} right shows the normalized average waveform for alphas and muons selected by F$_\text{prompt}$ for the EJ-286 and FB118 independent tests\footnote{When acquiring muons, the $\alpha$-source was placed in the highest part of the test chamber, to decrease the $\alpha$-particles rate above the threshold of 650 ADC channels}. The respective normalized waveforms overlap perfectly, showing that the same liquid argon purity is achieved in the two independent liquefaction run: such reproducibility is a non trivial achievement as the triplet half-life, hence the slope of the waveform tail strongly depend on the residual (N$_2$,O$_2$) impurity concentrations. 

\begin{figure}[htbp]
    \centering
    \includegraphics[width=.9\textwidth]{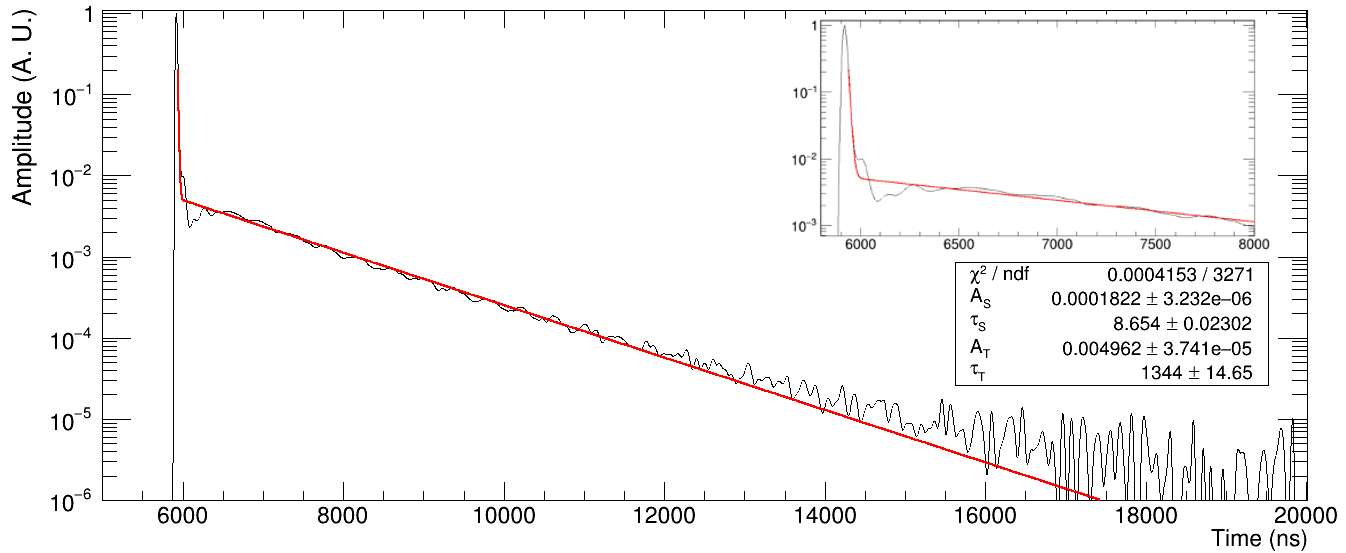}
    \includegraphics[width=.9\textwidth]{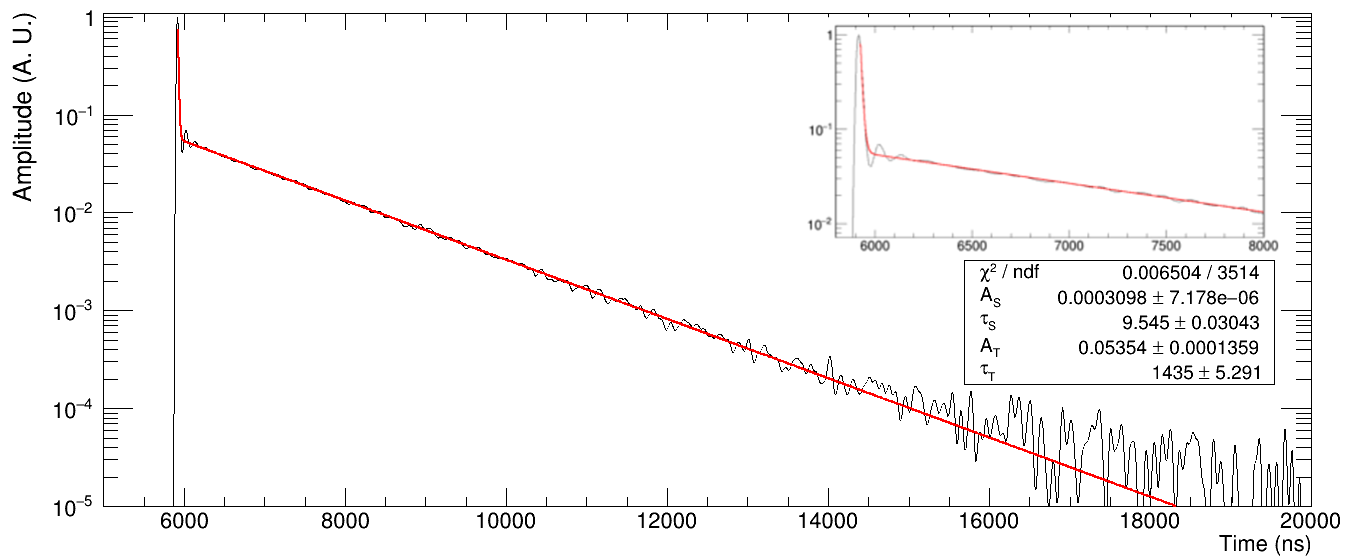}
    \caption{The deconvolved averaged waveforms of $\alpha$-particles (top) and muons (bottom). A fit of two exponential is made to retrieve the triplet time constant ($\tau_T$). The same fit is displayed with a zoom on the top right corner.}
    \label{fig:deconvolved_waveforms}
\end{figure}

To properly evaluate the triplet time constant, the average single p.e.\ waveform of Fig.~\ref{fig:sphe_calibration}, described by three exponentials convoluted with a Gaussian, is deconvolved from the averaged waveforms of Fig.~\ref{fig:averaged_waveforms} right: the Gold deconvolution algorithm of the TSpectrum library of Root Cern~\cite{root_cern} is used. The deconvolved waveforms shown in Fig.~\ref{fig:deconvolved_waveforms} for alphas (top) and muons (bottom), are fitted by the function $I(t)$  \cite{LAr_fund_properties}:
\begin{equation}
    \label{eq:fast_and_slow_fit}
    I(t) = A_S \exp{\left(-\frac{t}{\tau_S}\right)} + A_T \exp{\left(-\frac{t}{\tau_T}\right)},
\end{equation}
where $A_S$ and $A_T$ are the relative amplitudes and $\tau_S$ and $\tau_T$ are the time constants of the singlet and of the triplet dimer states respectively.

The $\tau_S$ and $\tau_T$ were computed by the fit of Eq.~\ref{eq:fast_and_slow_fit}. $A_S$ was taken by integrating the deconvolved waveform of Fig.~\ref{fig:deconvolved_waveforms} from the rising edge up to 4 times $\tau_S$ and the $A_T$ by integrating the triplet component fitted from the rising edge up to 20~$\mu$s. 
In Table~\ref{tab:lar_parameters}, for each measurement (row) we report the average values among all the available runs. The uncertainties are the combination of the individual standard deviations with the uncertainty of the average. 
The ratio $\langle A_S/A_T \rangle$ was computed from all the runs within each measurement, the ratio weighted average is then computed, properly propagating the uncertainties.
As shown in Table \ref{tab:lar_parameters}, we found consistent values among the three measurements both for the singlet/triplet decay times and for the relative amplitudes, hence we are allowed to average them: hence we quote $1414\pm21$~ns and $1294\pm35$~ns for the the average triplet decay time constant for muons and alphas respectively. The uncertainties include both the statistical and the systematics which are dominated by the single p.e.\ calibration.

\begin{table}[htbp]
\centering
 \caption {The singlet and triplet decay time constants ($\tau_S$ and $\tau_T$) and their relative amplitudes ($A_S$ and $A_T$) from the waveforms fit for each liquefaction and their averaged values.}
\label{tab:lar_parameters}
\smallskip
\begin{tabular}{|c|c|c|c|c|c|c|}
\hline
                  & & $\tau_S$ (ns) & $\tau_T$ (ns) & $A_S$ & $A_T$ & $A_S/A_T$       \\ \hline
\multirow{3}{*}{$\alpha$} & EJ-286 w.o.\ Vikuiti & 9.3~$\pm$~0.9 & 1287 $\pm$ 35 & 0.76~$\pm$~0.06 & 0.24~$\pm$~0.03 & 3.10 $\pm$ 0.42 \\ 
& EJ-286 w.\ Vikuiti &  11.9~$\pm$~1.8 & 1286~$\pm$~37 & 0.77~$\pm$~0.06 & 0.23~$\pm$~0.03 & 3.41 $\pm$ 0.51 \\ 
& FB118 w.\ Vikuiti & 7.8~$\pm$~1.8 & 1331~$\pm$~22 & 0.79~$\pm$~0.07 & 0.21~$\pm$~0.03 & 3.67 $\pm$ 0.58 \\  \hline
& Average & 9.7~$\pm$~3.0 & 1294 $\pm$ 35 & 0.77~$\pm$~0.04 & 0.23~$\pm$~0.02 & 3.35 $\pm$ 0.28 \\ \hline\hline
\multirow{3}{*}{$\mu$} & EJ-286 w.o.\ Vikuiti & 10.6~$\pm$~0.5 & 1371 $\pm$ 18 & 0.22~$\pm$~0.04 & 0.78~$\pm$~0.10 & 0.28 $\pm$ 0.07 \\
& EJ-286 w.\ Vikuiti & 9.6~$\pm$~0.1 & 1411 $\pm$ 33 & 0.23~$\pm$~0.05 & 0.77~$\pm$~0.11 & 0.29~$\pm$~0.08 \\
& FB118 w.\ Vikuiti & 10.6~$\pm$~4.4 & 1459 $\pm$ 35 & 0.23~$\pm$~0.05 & 0.77~$\pm$~0.11 & 0.30~$\pm$~0.08 \\ \hline
& Average & 10.2~$\pm$~5.1              & 1414 $\pm$ 21 & 0.18~$\pm$~0.03 & 0.82~$\pm$~0.04 & 0.29 $\pm$ 0.03 \\ \hline
\end{tabular}
\end{table}

This assesses that the overall quality of the chamber preparation and GAr liquefaction procedures are under control, and allows the direct comparison of the FB118 to the EJ-286 $\alpha$-peak structure.

\begin{figure}[htbp]
\center
\includegraphics[width=0.25\textwidth]{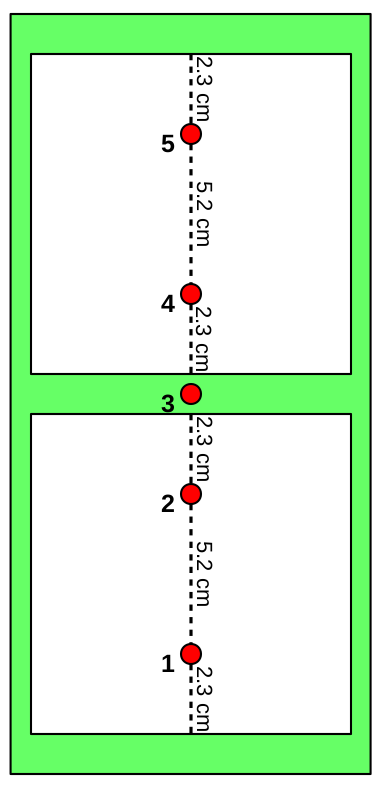}
\caption{The five height positions at fixed distance from the X-Arapuca used to characterize the WLS bars.}
\label{fig:points_scheme}
\end{figure}

The alpha source was fixed on the rail at a distance of $5.5\pm0.2$~cm from the dicroic filters (see Fig.~\ref{fig:cryosetup}) and was moved to five different positions to scan the DUT as shown in Fig.~\ref{fig:points_scheme}. One position is at the center of the optical cell and the other four are symmetrical to it, two for each window.

\begin{figure}[htbp]
    \centering
    \includegraphics[width=0.456\textwidth]{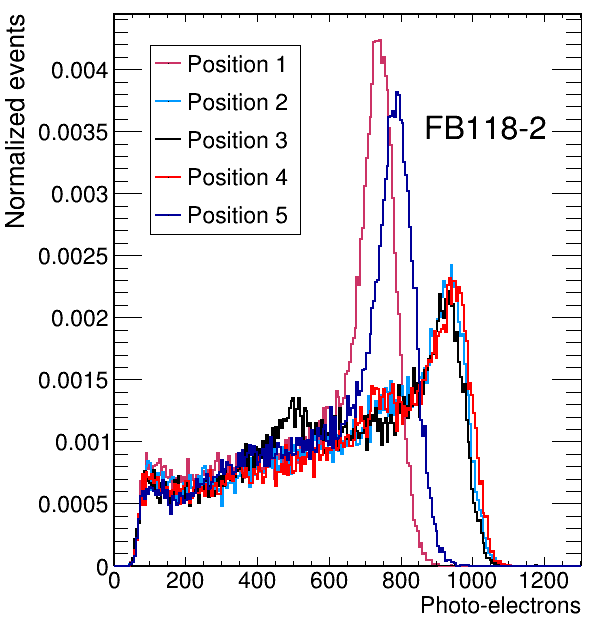}
    \includegraphics[width=0.45\textwidth]{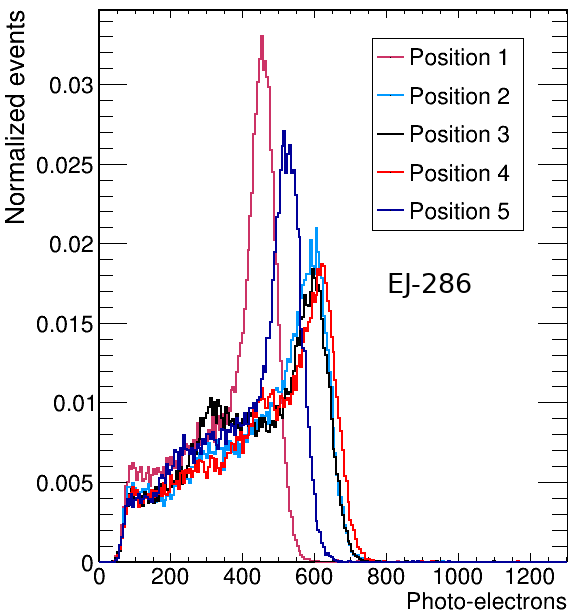}
    \caption{The $\alpha$ spectrum in number of detected photoelectrons  for each of the five source positions: the X-Arapuca is equipped with FB118 (left) and EJ-286 (right) respectively.}
    \label{fig:alpha_spectrums_lar}
\end{figure}

For each of the five source positions, the waveforms are integrated from 5.6 to 11~$\mu s$ and individually calibrated in term of s.p.e.\ as discussed above. Figure~\ref{fig:alpha_spectrums_lar} shows the calibrated $\alpha$ spectrum, for both the FB118 (left) and EJ-286 (right). As expected from the geometry, the spectra for positions 2, 3 and 4 overlaps, while a significant difference is found between position 1 and 5, expected to be identical: this is attributed to the shadowing of the $\alpha$-source holder that penalizes position 1 which, for this reason, is not used for the X-Arapuca efficiency ($\epsilon$) evaluation.

For each position a sizeable increase of the collected photons is observed when the optical cell is equipped with the FB118 instead of the EJ-286.

\begin{figure}[htbp]
    \centering
    \includegraphics[width=0.9\textwidth]{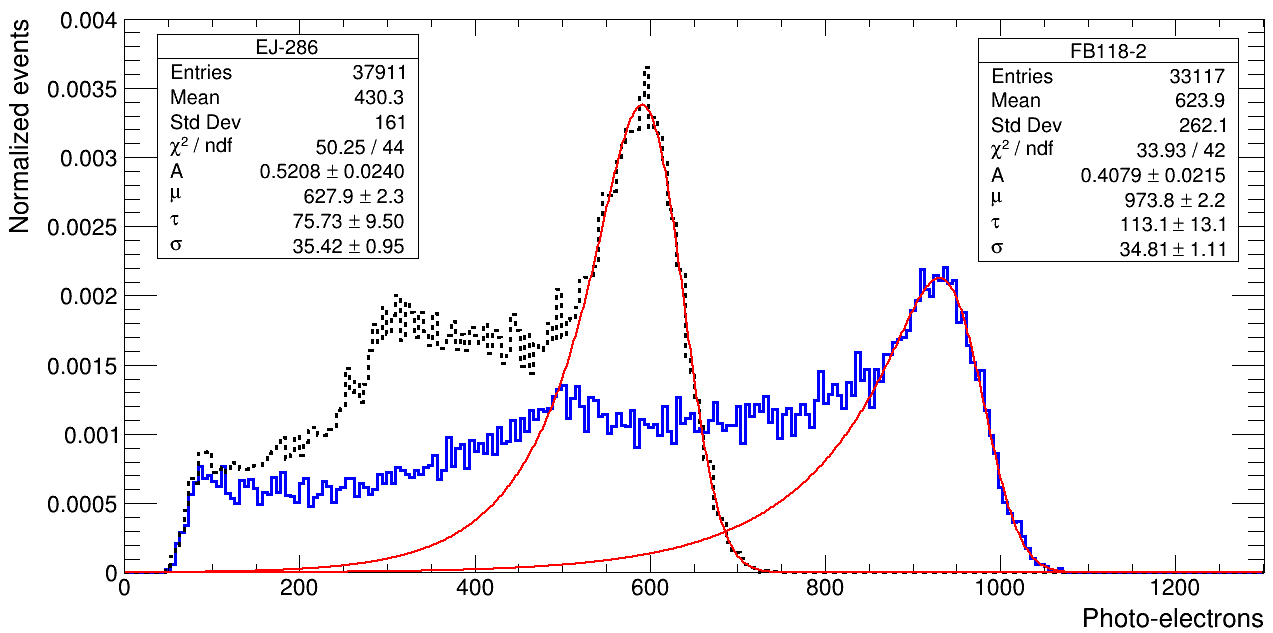}
    \caption{The $\alpha$ spectrum fit with Eq.~\ref{eq:alpha_spectrum} for the source in the central position (\#3)  for EJ-286 (black, dashed line) and FB118 (blue, solid line).}
    \label{fig:sample_fits_spectrum_lar}
\end{figure}

To determine the alpha peak position, we fit the spectrum with the convolution of a Gaussian and an exponential~\cite{alpha_equation}:
\begin{equation}
\label{eq:alpha_spectrum}
F(E) = \frac{A}{2\tau}\exp{\left(\frac{E-\mu}{\tau} + \frac{\sigma^2}{2\tau^2}\right)}\;\erf{\left(\frac{1}{\sqrt{2}}\left(\frac{E-\mu}{\sigma} +\frac{\sigma}{\tau}\right)\right)},
\end{equation}
where $\mu$ is the peak position, A is the area, $\sigma$ is the spread of the alpha-peak and $\tau$ is the slope of the tail on the peak low-energy side. Figure~\ref{fig:sample_fits_spectrum_lar} shows the spectrum  when the source is in the central position (\#3), and superimposed the fits for both the EJ-286 and FB118.

To estimate the X-Arapuca efficiencies, 
the scintillation light generated by the $\alpha$ source is assumed to be point like with a light yield of 35,000 photons/MeV (assuming the $\alpha$ quenching factor of 0.7)~\cite{LAr_fund_properties,model_nuclear_recoil_nl} and the number of photons reaching the X-Arapuca entrance window is numerically computed for each of the 5 source positions with the source-to-device distance of 5.5~$\pm$~0.2~centimeters. Table \ref{tab:expected_values_mc} shows the expected number of photoelectrons for a 1\% efficiency X-Arapuca device, and the detected ones for the three  tested WLS slab configurations. As discussed, position 1 is not considered.

\begin{table}[htbp]
\centering
\caption{Expected value of photons reaching the X-Arapuca acceptance window given by numerical evaluation described and number of photoelectrons detected in the three different configurations. An average of position 2, 3 and 4 was taken.}
\label{tab:expected_values_mc}
\smallskip
\begin{tabular}{|c|c|c|c|c|}
\hline
Positions & Expected (MC)               & EJ-286 w/o Vikuiti\textsuperscript{\textregistered} & EJ-286 w/ Vikuiti\textsuperscript{\textregistered} & FB118      \\ 
    & \# p.e & \# p.e.\ & \# p.e.\ & \# p.e.\ \\ \hline
2,3,4     & 288 $\pm$ 6  & 607 $\pm$ 9           & 637 $\pm$ 9          & 986 $\pm$ 14 \\ \hline
5         & 236 $\pm$ 5  & 515 $\pm$ 13          & 548 $\pm$ 14         & 825 $\pm$ 20 \\ \hline
1         & 236 $\pm$ 5 & 443 $\pm$ 11          & 473 $\pm$ 12         & 770 $\pm$ 18 \\ \hline
\end{tabular}
\end{table}

Finally a dedicated measurement to evaluate the SiPMs cross-talk was performed. The SiPMs were submerged in liquid nitrogen at dark and 1000 events were continuously recorded while triggering at 0.5 p.e.. The cross-talk probability is defined as the ratio between the number of events above 1 p.e.\ and the total number of the events~\cite{SiPM_better}. We measured a cross-talk probability ranging from 20~to~24\%,  corresponding to  1.22~$\pm$~0.02 avalanches/photon. Afterpulses, i.e. the number of secondary pulses within 10 $\mu$s contribute for 11.7\%.

\section{Results}
\label{sec:result}
%
To compute the absolute efficiency ($\epsilon$) of the device under test, two correction factors are applied: 1. re-normalization of the detected to expected  (for zero impurity concentrations) photon number, i.e. correction for the actual LAr purity monitored in each run; 2. correction for SiPMs cross-talk.
\begin{enumerate}
    \item The re-normalization factor (P$_{\text{LAr}}$) accounts for the fraction of the triplet component lost by impurities in LAr:
    \begin{equation}
        \label{eq:lar_purity_correction}
        P_{\text{LAr}} = \left(0.77+0.23\times\frac{\tau_{T}}{1414 \text{ ns}}\right)^{-1},
    \end{equation}
    where $\tau_T = 1294 \pm 35$~ns is the measured triplet time constant. Assuming
    the singlet and triplet contributions to be 0.77 and 0.23 and the theoretical triplet time constant of 1414~ns~\cite{first_lar_test,nitrogen_contamination_roberto,abudance_dependence}, P$_{LAr}$ maximum values is +2.6\%.
    \item The measured cross-talk probability leads to a correction of $-(18\pm1\%)$ 
    of the calibrated alpha spectrum.
\end{enumerate}
 
 These corrections are not applied in the relative comparison of the two WLS slabs, as their contribution evens out.
 
The measured absolute efficiencies of the X-Arapuca with EJ-286 (with and without Vikuiti\textsuperscript{\textregistered}) and FB118 with Vikuiti\textsuperscript{\textregistered} are displayed in Table~\ref{tab:lar_results_efficiency}. The efficiency uncertainty includes those from LAr purity correction,  the single photo-electron calibration and the solid angle. Table~\ref{tab:lar_results_efficiency} reports also the signal-to-noise ratio (defined as $\mu_1/\sqrt{\sigma_0^2 + \sigma_1^2}$, where $\mu_1$ is the mean charge of a single phe, $\sigma_0$ and $\sigma_1$ are the standard deviation of the noise and single phe, respectively, the Gain (Fig.~\ref{fig:sphe_calibration}) when the source is at position 3 and the resolution ($\sigma/\mu$) of the alpha peak ($\sigma \text{ and } \mu$ from the fit function~\ref{eq:alpha_spectrum}): the energy resolution scales as expected with the number of detected photons.
\begin{table}[htbp]
\centering
\caption{Summary of the relevant facts for the X-Arapuca efficiency measurements by $^{241}$Am $\alpha$ source radiation  and the three considered configurations: SPE gain, as defined in section~\ref{sec:measurements_lar}, Energy Resolution of the $\alpha$ peak, S/N: Signal to Noise, $\epsilon_{\text{raw}}$, $\epsilon$; Efficiency Prior (raw) and post corrections  respectively. $\tau_T$:~ measured Triplet half-life}
\label{tab:lar_results_efficiency}
\smallskip
\begin{tabular}{|c|c|c|c|}
\hline
                    & EJ-286 w/o Vikuiti & EJ-286 w/ Vikuiti & FB118           \\ \hline
SPE Gain (ADC$\cdot$ns) & 1680~$\pm$~80               & 1690~$\pm$~80              & 1735~$\pm$~90          \\ \hline
En. res. ($\sigma/\mu$)     & 6.3 $\pm$ 0.2 \%     & 6.0 $\pm$ 0.2 \%
    & 3.6 $\pm$ 0.1 \%\\ \hline
S/N                 & 6.8~$\pm$~0.3                   & 7.3~$\pm$~0.3                  & 7.3~$\pm$~0.3               \\ \hline
$\epsilon_\text{raw}$         & 2.1 $\pm$ 0.1 \%      & 2.3 $\pm$ 0.1 \%     & 3.5 $\pm$ 0.1 \%  \\ \hline\hline
$\tau_T$           & \multicolumn{3}{c|}{1294~$\pm$~35~ns}                                   \\ \hline
LAr purity correction          & \multicolumn{3}{c|}{+ (1.4 to 2.6)~\%}                                   \\ \hline
Cross-talk correction           & \multicolumn{3}{c|}{- (18~$\pm$~1)~\%}                                   \\ \hline
$\epsilon$         &  1.8 $\pm$ 0.1\%      &  1.9 $\pm$ 0.1\%     &  2.9 $\pm$ 0.1\%  \\ \hline
\end{tabular}
\end{table}

From the ratio of the alpha-peak means, for each source position, we compute the PDE variation $G_\epsilon$ when the X-Arapuca has the FB118 as secondary wavelength shifter instead of the EJ-286 Table~\ref{tab:lar_results_relative}, defining $G_\epsilon$ as:
\begin{equation}
G_\epsilon = \left(\frac{\mu_{\text{FB118}}}{\mu_{\text{EJ-286}}}-1\right)\times 100 \; [\%],
\end{equation}
where $\mu$ is the alpha-peak mean from Eq.~\ref{eq:alpha_spectrum} averaged over positions 2, 3 and 4. The 2.5\% uncertainty on the peak position ($\mu$) is driven the gain calibration uncertainty.
\begin{table}[htbp]
\centering
\caption{\label{tab:lar_results_relative} The PDE increase ($G_\epsilon$) when replacing the EJ-286 in the X-Arapuca with the FB118, for the five source positions as in Table~\ref{tab:expected_values_mc}. The final $G_\epsilon$ we quote is the average of positions 2, 3 and 4.}
\smallskip
\begin{tabular}{|c|c|}
\hline
Positions & $G_\epsilon$  \\ \hline
2,3,4     & 55 $\pm$ 5 \%  \\ \hline
5         & 50 $\pm$ 5 \%  \\ \hline
1         & 63 $\pm$ 6 \%   \\ \hline
\end{tabular}
\end{table}

When the Vikuiti\textsuperscript{\textregistered} is applied on the WLS slab edges the PDE increases from 3~to~11\% where the extreme source positions (1 and 5) show the largest increase as measured at room temperature (Sec.~\ref{sec:measurement_rt}). The increase of the PDE measured in LAr is lower than in air/vacuum, as in LAr  the critical angle is about 61$^\circ$ and 54$^\circ$ for FB118 and EJ-286 respectively, while in air it is 42$^\circ$  and 39$^\circ$, hence in LAr photons are less trapped in the FB118 light-guide than in the EJ-286 by total internal reflection at the slab surfaces.

Another relevant achievement of this work is the assessment of a clean and reproducible procedure to liquefy gAr, providing high LAr purity despite the sizeable payload in the chamber. This not only makes negligible the PDE correction for the LAr purity, but allows to average the singlet and triplet time constants and their relative fractions as reported in Table~\ref{tab:lar_parameters} for both alphas and muons. 
The parameters found in this work are in good agreement with LAr scintillation properties found in previous investigations \cite{nitrogen_contamination_roberto,LAr_fund_properties,protoDUNE_first_results}.
%

\section{Conclusions}
In this work we optimized and accurately measured, prior and after our actions, the PDE and PDU of the X-Arapuca: it is the basic unit of the photon detection system of the Dune Far Detector. In particular we investigated the impact on the performances, by changing the specular reflector configuration inside the device and the type of secondary WLS slab: we tested a new WLS material developed by us in collaboration with G2P Co.~(\cite{G2P}) for LAr applications and measured the perfomances of the X-Arapuca device when adopting it instead of the off-shell baseline product by Eljen Co.. With two dedicated setups, we performed a set of measurements in air at room temperature and both in vacuum and in LAr at cryogenic temperatures. At room temperature we found that the PDU is enhanced of $\sim$40~\% by applying the extended specular reflector on the WLS slab edges and that the PDE increases of $\sim$50~\% with the newly developed G2P WLS w.r.t. the Eljen one. In LAr, we readout for the first time the SiPMs by a  trans-impedance amplifier and derived the absolute and relative PDEs, of the X-Arapuca in three different configurations by scanning with an $^{241}$Am $\alpha$ source: we found that the PDE is enhanced by $\sim$~50\% with the G2P WLS w.r.t. the baseline Eljen product, and about 10~$\pm$~1\%  when applying the reflector on the slab edges. The $\alpha$ scanning confirms that the PDU is enhanced in this configuration.
As for the absolute PDEs we found (1.8~$\pm$~0.1)\% for the standard device configuration, and (2.9~$\pm$~0.1)\% when we substitute the Eljen with the G2P WLS and the reflector on its edges. These values are net from SiPMs cross-talk. The PDE of (1.8~$\pm$~0.1)\% that we measure for the standard device configuration, is in tension with the previous measurement \cite{first_lar_test} of (3.5~$\pm$~0.5)\%: we point out that in the latter the SiPMs were operated with an overvoltage of +5 V (here +2.7 V), the S/N was poorer as no cold front-end stage was present, and the alpha source was in a fixed position in front of the (half-size) X-Arapuca device. 

The novel WLS material that we have developed has a wide potential in LAr based experiments as LEGEND\cite{Legend} and Dark Side\cite{darksidecollaboration2021separating} or SBND\cite{sbnd}

As for LAr purity we showed that our procedures allow to attain high purity with high reproducibility. This allows the direct comparison of PDEs from different LAr liquefaction processes as negligible corrections and associated uncertainties are required. 
The high LAr purity allow to measure the main Ar$^*$ dimers de-excitation parameters, namely the singlet and triplet decay time and their relative amplitudes for both $\alpha$ particles and muons: we found them in agreement with \cite{first_lar_test,nitrogen_contamination_roberto}.

\acknowledgments
This work was funded by INFN GERDA/LEGEND and DUNE projects, and in part by the Coordena\c{c}\~ao de Aperfei\c{c}oamento de Pessoal de N\'ivel Superior - Brasil (CAPES) - Finance Code 001

\bibliographystyle{JHEP}
\bibliography{references}

\end{document}